\documentstyle[epsfig,aps,12pt]{revtex}
\begin{document}
\renewcommand{\thefigure}{\arabic{figure}}
\title{Effective interactions between parallel-spin electrons in two-dimensional jellium approaching the magnetic phase transition}
\author{B. Davoudi$^{1,2}$, M. Polini$^1$, G. Sica$^1$, M. P. Tosi$^{1}\footnote{Corresponding author; e-mail:tosim@sns.it}$}
\address{
$^1$NEST-INFM and Classe di Scienze, Scuola Normale Superiore, I-56126 Pisa, Italy\\ 
$^2$Institute for Studies in Theoretical Physics and Mathematics, Tehran, P.O.Box 19395-5531,Iran\\
}
\maketitle
\vspace{1cm}
\hspace{-0.7cm}{\Large Abstract}\\

We evaluate the effective interactions in a fluid of electrons moving in a plane, on the 
approach to the quantum phase transition from the paramagnetic to the fully spin-polarized 
phase that has been reported from Quantum Monte Carlo runs. We use the approach of 
Kukkonen and Overhauser to treat exchange and correlations under close constraints imposed 
by sum rules. We show that, as the paramagnetic fluid approaches the phase transition, the 
effective interactions at low momenta develop an attractive region between parallel-spin electrons 
and a corresponding repulsive region for antiparallel-spin electron pairs. A connection with the 
Hubbard model is made and used to estimate the magnetic energy gap and hence the temperature 
at which the phase transition may become observable with varying electron density in a 
semiconductor quantum well.

\vspace{1cm}
\hspace{-0.7cm}{\it Keywords:} A. Quantum wells; D. Electron-electron interactions; D. Phase transitions.

\hspace{-0.7cm}PACS numbers: 05.30.Fk, 71.10.Ca\\

\newpage

Electronic fluids with an essentially two-dimensional (2D) dynamics, as are realized in 
semiconductor heterostructures, present a very rich phenomenology especially at low density, 
where the interactions between carriers come to play a crucial role [1]. This fact continues to 
motivate theoretical studies of the simplified system consisting of a gas of electrons described 
by the basic Coulomb Hamiltonian with a uniform neutralizing background (EG). More 
generally, this has been for many decades a basic reference model for calculations of electronic 
structure in molecular and condensed-matter systems [2].\\

	Quantum Monte Carlo (QMC) simulation studies have been highlighting the formation of 
spontaneous spin polarization in the EG at strong coupling. A continuous transition from the 
paramagnetic to spin-polarized states has been reported to take place in the 3D EG with 
increasing coupling strength [3, 4], before a first-order transition into a ferromagnetic Wigner 
electron crystal occurs. Spontaneous partial spin polarization ("weak ferromagnetism") has been 
observed in doped hexaborides [5]. In the 2D EG, on the other hand, the quantum simulation 
studies indicate a first-order transition to a fully spin-polarized ("ferromagnetic") fluid before 
crystallization, no evidence being found for partially spin-polarized states [6 - 8].\\

	Theoretical studies of the equation of state described by the ground-state energy $E(n,\zeta)$ of 
the EG with spin densities $n_\uparrow$ and $n_\downarrow$ as a function of the electron density $n=n_\uparrow+n_\downarrow$ and of the spin polarization  $\zeta=(n_\uparrow-n_\downarrow)/n$ are clearly called for in both dimensionalities. The ground-state energy is in turn determined by the pair distribution functions, stressing the importance of two-body electron-electron scattering via effective interactions in the many-body system [9, 10]. \\

	Exchange between parallel-spin electrons and correlations from the Coulomb repulsions are 
crucial in determining the pair distribution functions. These effects are embodied in the so-called 
local-field factors (LFF) entering the expressions of the charge and spin susceptibilities of the 
EG [11]. The LFF play the same role as vertex corrections in accounting for the difference 
between the effective potentials experienced by an electron and their mean-field value. From their 
knowledge one can also calculate the quasi-particle self-energies and the effective electron-
electron interactions, as is well known from work on the unpolarized EG [12].\\

	In this letter we are concerned with the effective electron-electron interactions in the 
paramagnetic phase of the 2D EG as functions of the coupling strength on the approach to the 
magnetic quantum phase transition. We adopt the linear-response approach of Kukkonen and 
Overhauser [9] to construct these interactions from recent determinations of the LFF [13, 14], 
embodying their asymptotic behaviours from exact sum rules and information from QMC 
studies. The difference between the $\uparrow\uparrow$ and $\uparrow\downarrow$ effective interactions is related to the magnetic susceptibility and in the low scattering-momentum region most strikingly reflects the 
approaching phase transition in its behaviour with increasing coupling strength. After pointing 
out how these results could be used to reveal the magnetic phase transition in a theoretical 
calculation based on the solution of Schr\"odinger equations for two-electron problems, we 
establish an approximate connection with a simple mean-field solution of the Hubbard model for 
itinerant electrons and use it to estimate the temperature of the magnetic transition in the 2D EG 
at the density predicted by the QMC studies.\\

	Following, therefore, Kukkonen and Overhauser [9], the effective interaction potential 
between two electrons with spin indices $\sigma_1$ and $\sigma_2$ is written in momentum space as
\begin{equation}
V_{\sigma_1\sigma_2}(q)=V_0(q)-J(q)\sigma_1\cdot\sigma_2
\end{equation}
where by convention the product $\sigma_1\cdot\sigma_2$ equals +1 for parallel spins and -1 for antiparallel spins. The functions $V_0(q)$ and $J(q)$ are expressed in terms of the bare Coulomb potential $v(q)=2\pi e^2/q$ and of the ideal-gas Lindhard susceptibility $\chi_0(q)$ as
\begin{equation}
V_0(q)=v(q)\frac{1+[1-G_+(q)]G_+(q)v(q)\chi_0(q)}{1+[1-G_+(q)]v(q)\chi_0(q)}
\end{equation}
and
\begin{equation}
J(q)=v(q)\frac{G_-^2(q)v(q)\chi_0(q)}{1-G_-(q)v(q)\chi_0(q)}
\end{equation}

Here, $G_+(q)$ and $G_-(q)$ are the charge-charge and spin-spin LFF in the paramagnetic 2D EG. It 
is important to remark for later use that the denominator in Eq. (3) also enters the microscopic 
magnetic susceptibility $\chi(q)$ of the EG, which is given by
\begin{equation}
\chi(q)=\frac{\chi_0(q)}{1-G_-(q)v(q)\chi_0(q)}
\end{equation}
in units such that the Bohr magneton equals unity.\\

	The limiting behaviours of the LFF at low and high momenta are determined by a set of exact 
sum rules [15, 16]. Considering first $G_+(q)$, these are the compressibility sum rule yielding
\begin{equation}
\lim_{q\rightarrow 0}G_+(q)=\left\{\frac{1}{\pi}-\frac{r_s^3}{8\sqrt{2}}\left[\frac{d^2\varepsilon_c(r_s)}{dr_s^2}-\frac{1}{r_s}\frac{d\varepsilon_c(r_s)}{dr_s}\right]\right\}\frac{q}{k_F}
\end{equation}
with the Fermi momentum given by $k_F=\sqrt{2\pi n}$ in terms of the areal density $n$ , and the Kimball-Niklasson-Holas relation yielding
\begin{equation}
\lim_{q\rightarrow\infty} G_+(q)=[1-g(0)]-\frac{r_s}{2\sqrt{2}}\frac{d[r_s\varepsilon_c(r_s)]}{dr_s}\frac{q}{k_F}.
\end{equation}		                            
In these equations $\varepsilon_c(r_s)$ is the correlation energy per particle as a function of the density parameter       $r_s=(\pi na_B^2)^{-1/2}$ with $a_B$ the effective Bohr radius, and $g(0)$ is the pair distribution 
function at contact. The corresponding limiting values of $G_-(q)$ are determined by the magnetic 
susceptibility sum rule,
\begin{equation}
\lim_{q\rightarrow 0}G_-(q)=\left[\left.\frac{1}{\pi}-\frac{r_s}{2\sqrt{2}}\frac{\partial^2\varepsilon_c(r_s)}{\partial \zeta^2}\right|_{\zeta=0}\right]\frac{q}{k_F}
\end{equation}
and by the Santoro-Giuliani-Holas relation,
\begin{equation}
\lim_{q\rightarrow\infty} G_-(q)=g(0)-\frac{r_s}{2\sqrt{2}}\frac{d[r_s\varepsilon_c(r_s)]}{dr_s}\frac{q}{k_F}.
\end{equation}		                            
Ample data from QMC studies are available for the correlation energy of the 2D EG as a 
function of density [7, 17] and for its long-wavelength magnetic susceptibility determining the 
coefficient in Eq. (7) [8, 18 - 20]. The value of $g(0)$ is also known over a wide range of density 
[21], from the solution of the two-body scattering problem and from many-body calculations in 
the ladder approximation.\\

	Complete analytical expressions for both LFF have been obtained in the density range $r_s\leq 10$[13, 14], using all the above information and further QMC data from [18 - 20] to describe 
their behaviour in the region of intermediate wave number. We have now extended this work to 
determine the LFF at $r_s= 20$, with the results that are shown in Figure 1. Especially relevant for 
this purpose have been the newly available QMC data on the magnetic susceptibility [8], which 
reaches the value $\chi(0)/\chi_0(0)\cong 25$ at  $r_s= 20$. The corresponding values of the effective 
interactions between parallel-spin and antiparallel-spin electrons at several values of $r_s$ are 
shown in Figures 2 and 3, respectively. Some smoothing of apparently noisy features has been 
made in obtaining these results.\\

	As was shown in Ref. [13, 14] and as again assumed in constructing $G_-(q)$ in Figure 1, the 
long-wavelength behaviour reported for this LFF in Eq. (7) holds over a rather wide range of 
scattering momentum $q$, extending almost up to $q=2k_F$. The plateau shown by the effective 
interactions in this range of momentum is, therefore, essentially determined by the value of the 
magnetic susceptibility $\chi(0)$. As already noted, this thermodynamic parameter is approaching 
very large values as the paramagnetic phase of the 2D EG approaches the quantum phase 
transition to a fully spin-polarized state, that the latest QMC evidence places at $r_s\approx 25$[8]. The 
effective interaction between parallel-spin electrons in Figure 2 becomes correspondingly 
strongly attractive in the momentum range $q\leq 2k_F$, while a repulsive structure grows at 
momenta above $2k_F$. In parallel to this behaviour, and as is evident from Eq. (1), the effective 
interaction between antiparallel-spin electrons in Figure 3 becomes strongly repulsive in the 
momentum range $q\leq 2k_F$. Although the quantitative details of these results are rather sensitive 
to the precise value of  $\chi(0)$, their significance is evidently correct: the approach to the transition 
into an ordered phase having a space-independent order parameter is driven by the emergence of 
effective interactions favouring spin alignments over a scale of distances which is becoming long 
compared with $(2k_F)^{-1}$.\\

	In Figure 2 we have also reported the values of $V_{\uparrow\uparrow}(q)$ as calculated in the fully spin-
polarized state at $r_s=40$[22], to show that the effective interactions regain a "standard shape" 
after the magnetic phase transition has taken place. That is, the electron-electron interactions in 
momentum space are repulsive and show a rapid variation as $q$ goes through $2k_F$, corresponding 
to Friedel oscillations in coordinate space.\\

	As was shown in the work of Takada [23] (see also Richardson and Ashcroft [24]), the 
effective electron-electron interactions may be used to assess the possibility of superconductive 
s-wave or p-wave pairing in the EG through the solution of an Eliashberg equation in the 
absence of phonons. It is natural to ask whether they may play a similarly useful role in regard 
to a magnetic phase transition. In fact, Ortiz et al. have stressed in their work [4] that a precise 
signal of the transition of an electron fluid to a magnetically ordered state is carried by the spin-
spin radial distribution function, which is proportional to the difference $g_{\uparrow\uparrow}(r)-g_{\uparrow\downarrow}(r)$ between the radial distribution functions for the two spin populations. Near a magnetic transition the function $g_{\uparrow\uparrow}(r)-g_{\uparrow\downarrow}(r)$ can be expected to change sign at short range, going from negative in 
the deep paramagnetic phase (where it reflects a preference for spin alternation) to positive as 
magnetically ordered domains emerge. Methods to evaluate pair distribution functions in a 
many-body system through the solution of a Schr\"odinger equation for a two-body scattering 
problems have drawn some attention in the recent literature [10, 21, 25] and could easily be 
adapted to the study of magnetic phase transitions with the help of the effective potentials that we 
have determined in this work.\\

	Here we adopt a much simpler and cruder route in order to estimate the magnetic energy gap $\Delta$  which would be required for spin flips at zero momentum transfer in the magnetically ordered phase. This estimate is based on the remark that Eq. (4) for the magnetic susceptibility takes at low momenta the form that is obtained in the Hubbard model for a system of itinerant electrons [26]. The value of the Hubbard parameter $U$ appropriate for the 2D EG near the magnetic transition can therefore be estimated as
\begin{equation}
U\approx \lim_{k\rightarrow 0}\left[G_-(k)v(k)\right]=\chi_0^{-1}(0)-\chi^{-1}(0)
\end{equation}
Setting $\Delta\approx nU$ for a fully spin-polarized state we estimate
\begin{equation}
\Delta\approx k_BT_c\approx\frac{2}{r_s^2}\left(1-\frac{\chi_0(0)}{\chi(0)}\right) {\rm Ryd},
\end{equation}
that is, taking $r_s\approx 25$ and $\chi(0)/\chi_0(0)\approx 25$ as from Ref. [8], $\Delta\approx 3.10^{-3}\;{\rm Ryd}$ and $T_c\approx  400\; {\rm K}$ for the EG model. This value is reduced by about three orders of magnitude when one takes account of an effective mass correction and of screening by a background dielectric constant in a 
semiconducting material. For instance, for the high-mobility GaAs/GaAlAs heterostructures 
studied by Yoon et al.[27] we estimate $T_c\approx 1\; {\rm K}$.\\

	In summary, we have evaluated the effective interactions in momentum space between 
electrons in the paramagnetic phase of the 2D EG at various densities and shown how their 
shape reflects the emergence of magnetically ordered domains on the approach to the quantum 
phase transition into a fully spin-polarized phase. We have also given an estimate for the 
transition temperature at the density predicted by the QMC runs.\\

	This work was partially funded by MIUR under the PRIN2001 Initiative.\\
\newpage	
\vspace{1cm}
\hspace{-0.7cm}{\Large References}\\
\vspace{0.5cm}

[1] See, e.g. E. Abrahams, S. V. Kravchenko, M. P. Sarachik, Rev. Mod. Phys. 

\hspace{0.4cm} 73 (2001) 251.

[2] See, e.g. D. Ceperley, Nature, 397 (1999) 386.

[3] D. M. Ceperley, B. J. Alder, Phys. Rev. Lett. 45 (1980) 566.

[4] G. Ortiz, M. Harris, P. Ballone, Phys. Rev. Lett. 82 (1999) 5317.

[5] 	D. P. Young, D. Hall, M. E. Torelli, Z. Fisk, J. L. Sarrao, J. D. Thompson, H.-R. 

\hspace{0.4cm} Ott, S. B. Oseroff, R. G. Goodrich, R. Zysler, Nature 397 (1999) 412.

[6]	B. Tanatar, D. M. Ceperley, Phys. Rev. B 39 (1989) 5005.

[7]	D. Varsano, S. Moroni, G. Senatore, Europhys. Lett. 53 (2001) 348.

[8]	C. Attaccalite, S. Moroni, P. Gori-Giorgi, G. B. Bachelet, cond-mat/0109492.

[9]	C. A. Kukkonen, A. W. Overhauser, Phys. Rev. B 20 (1979) 550.

[10] A. W. Overhauser, Can. J. Phys. 73 (1995) 683.

[11] See, e.g. K. S. Singwi, M. P. Tosi, Solid State Phys. 36 (1981) 177.

[12] S. Yarlagadda, G. F. Giuliani, Phys. Rev. B 61 (2000) 12556, and references given 

\hspace{0.55cm} there.

[13] B. Davoudi, M. Polini, G. F. Giuliani, M. P. Tosi, Phys. Rev. B 64 (2001) 153101.

[14] B. Davoudi, M. Polini, G. F. Giuliani, M. P. Tosi, Phys. Rev. B 64 (2001) 233110.

[15] G. E. Santoro, G. F. Giuliani, Phys. Rev. B 37 (1988) 4813.

[16] M. Polini, M. P. Tosi, Phys. Rev. B 63 (2001) 045118.

[17] F. Rapisarda, G. Senatore, Austr. J. Phys. 49 (1996) 161.

[18] S. Moroni, D. M. Ceperley, G. Senatore, Phys. Rev. Lett. 69 (1992) 1837.

[19] G. Senatore, S. Moroni, D. M. Ceperley, in Quantum Monte Carlo Methods 

\hspace{0.55cm} in Physics and Chemistry, ed. M. P.Nightingale and C. J. Umrigar (Kluwer, 

\hspace{0.55cm} Dordrecht 1999).

[20] G. Senatore, S. Moroni, D. Varsano, Solid State Commun. 119 (2001) 335.

[21] M. Polini, G. Sica, B. Davoudi, M. P. Tosi, J. Phys.: Condens. Matter 13, 3591 

\hspace{0.55cm} (2001).

[22] G. Sica, M. Polini, M. P. Tosi, Mod. Phys. Lett. B 15 (2001) 1053.

[23] Y. Takada, Phys. Rev. B 47 (1993) 5202.

[24] C. F. Richardson and N. W. Ashcroft, Phys. Rev. B 54 (1996) R764.

[25] P. Gori-Giorgi, J. P. Perdew, Phys. Rev. B 64 (2001) 155102.

[26] T. Izuyama, D. J. Kim, R. Kubo, J. Phys. Soc. Japan 18 (1963) 1025.

[27] J. Yoon, C. C. Li, D. Shahar, D. C. Tsui, M. Shayegan, Phys. Rev. Lett. 82 (1999)

\hspace{0.55cm} 1744.

\newpage	
\vspace{1cm}
\hspace{-0.7cm}{\Large Figure caption}\\
\vspace{0.5cm}

\begin{figure}
\centerline{\mbox{\psfig{figure=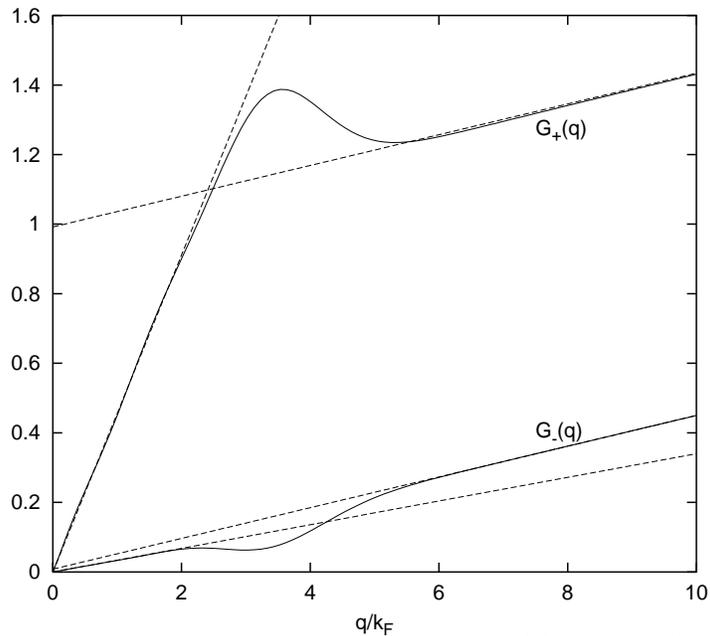, angle =0, width =10 cm}}} 
\caption{Charge-charge and spin-spin local field factors, $G_+(q)$ and $G_-(q)$, as functions of $q/k_F$ in the 2D EG in the paramagnetic state at $r_s=20$. The straight dashed lines show the asymptotic behaviours determined by sum rules and exact relations (Eqs. (5) - (8)), while the full line show an approximate interpolation suggested by earlier studies at lower $r_s$[13, 14].}
\label{Fig1}
\end{figure}

\begin{figure}
\centerline{\mbox{\psfig{figure=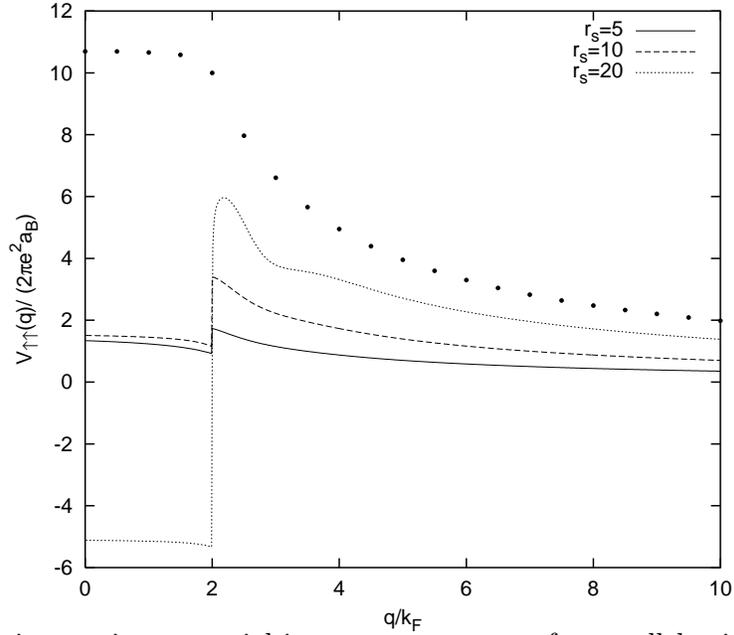, angle =0, width =10 cm}}} 
\caption{Effective interaction potential in momentum space for parallel-spin electrons in the 2D EG in the paramagnetic state at various values of $r_s$ from 5 to 20. The dots report the effective interaction in the fully spin-polarized 2D EG at $r_s= 40$[22].}
\label{Fig2}
\end{figure}

\begin{figure}
\centerline{\mbox{\psfig{figure=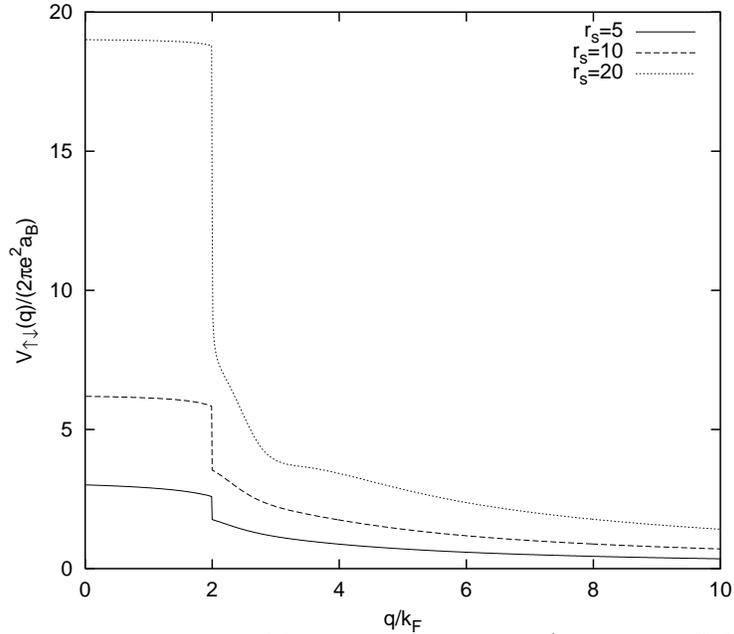, angle =0, width =10 cm}}} 
\caption{Effective interaction potential in momentum space for antiparallel-spin electrons in the 2D EG in the paramagnetic state at various values of $r_s$ from 5 to 20.}
\label{Fig3}
\end{figure}

\end{document}